\documentclass[prb,superscriptaddress,amsmath,amssymb,floatfix,showpacs,twocolumn]{revtex4-1}

\usepackage{amsfonts,amsmath,amssymb}
\usepackage{epsfig}
\usepackage{times}
\usepackage{graphicx}
\usepackage{bm}
\usepackage{xcolor}
\usepackage{dcolumn}
\usepackage{array}
\usepackage{float}

\newcommand\jeff{j_{\text{eff}}}
\newcommand\iridate{\text{Sr}_2\text{IrO}_4}
\newcommand\bk{{\mathbf{k}}}

\begin{document}

\title{Pseudogap and electronic structure of electron-doped $\iridate$}

\author{Alice Moutenet}
\affiliation{Centre de Physique Th\'eorique, \'Ecole Polytechnique, CNRS, Universit\'e Paris-Saclay, 91128 Palaiseau, France}
  \affiliation{Coll\`ege de France, 11 place Marcelin Berthelot, 75005 Paris, 
France}
\author{Antoine Georges} 
\affiliation{Coll\`ege de France, 11 place Marcelin Berthelot, 75005 Paris, France}
\affiliation{Center for Computational Quantum Physics, Flatiron Institute, 162 Fifth Avenue, New York, NY 10010, USA}
\affiliation{Centre de Physique Th\'eorique, \'Ecole Polytechnique, CNRS, Universit\'e Paris-Saclay, 91128 Palaiseau, France}
\affiliation{Department of Quantum Matter Physics, University of Geneva, 24 Quai Ernest-Ansermet, 1211 Geneva 4, Switzerland}
\author{Michel Ferrero}
\affiliation{Centre de Physique Th\'eorique, \'Ecole Polytechnique, CNRS, Universit\'e Paris-Saclay, 91128 Palaiseau, France}
  \affiliation{Coll\`ege de France, 11 place Marcelin Berthelot, 75005 Paris, France}

\date{\today}

\begin{abstract}
We present a theoretical investigation of the effects of correlations on the
electronic structure of  the Mott insulator $\iridate$ upon electron doping. A
rapid collapse of the Mott gap upon doping is found, and  the electronic
structure displays a strong momentum-space differentiation at low doping level:
The Fermi surface consists of pockets centered around $(\pi/2,\pi/2)$, while a
pseudogap opens near $(\pi,0)$. Its physical origin is shown to be related to
short-range spin correlations. The pseudogap closes upon increasing doping, but
a differentiated regime characterized by a modulation of the spectral intensity
along the Fermi surface persists to higher doping levels.  These results,
obtained within the cellular dynamical mean-field theory framework, are
discussed in comparison to recent photoemission experiments and an overall good
agreement is found. 
\end{abstract}

\maketitle

\newpage

\section{Introduction}

Understanding the 
physical mechanism responsible for the pseudogap in cuprate superconductors, and its interplay with superconductivity, 
is still a central and very debated question in the field of strongly-correlated materials. 
In this respect, the $\iridate$ iridate is a very interesting playground as it closely resembles these materials. 
It is indeed isostructural to $\text{La}_2\text{CuO}_4$,~\cite{randall_jacs_1957}
and its low-energy electronic structure is well described by a single half-filled band because of strong spin-orbit (SO) coupling 
as shown by electronic structure calculations in the Local Density Approximation (LDA).~\cite{kim_prl_2008, martins_prl_2011} 
Angular Resolved PhotoEmission Spectroscopy (ARPES), optical spectroscopy, and resonant inelastic X-ray 
scattering experiments~\cite{kim_prl_2008,jungho_kim_prl_2012,geneva_prl_2015,brouet_prb_2015,terashima_prb_2017} 
as well as scanning tunneling microscopy~\cite{nichols_prb_2014} demonstrate that 
the ground-state of this material is a Mott insulator.  
This conclusion is also supported by electronic structure calculations 
taking into account electronic correlations.~\cite{martins_prl_2011,zhang_prl_2013}
The similarity between the low-energy electronic structure of $\iridate$ and that of cuprates has led to the 
quest for superconductivity upon doping in this material.~\cite{watanabe_prl_2013, yang_prl_2014, kim_nature_2015, yan_prx_2015}

The electronic configuration of the $\text{Ir}^{4+}$ ions is $\{\text{Xe}\}f^{14}5d^5$ and $\iridate$ crystallizes in 
the $\text{K}_2\text{NiF}_4$ tetragonal structure, as $\text{La}_2\text{CuO}_4$ or $\text{Sr}_2\text{RhO}_4$.~\cite{randall_jacs_1957} The $\text{IrO}_6$ octahedra are rotated about the $c$-axis by $\sim 11$ deg., generating a doubled unit cell.~\cite{crawford_prb_1994}
The $5d^5$ electronic configuration would naively lead to a metallic state in a band theory approach. $\text{Sr}_2\text{RhO}_4$, 
having an identical atomic arrangement with nearly the same lattice constants and bond angles, is indeed found to be a Fermi liquid metal.~\cite{vogt_jssc_1996}
$\iridate$ however has a very strong SO coupling, a property which was shown to modify the electronic structure near the Fermi level in $5d$ systems.~\cite{singh_prb_2002,rossnager_prb_2006,xiang_prb_2007,martins_prl_2011,zhang_prl_2013}
This compound then effectively reduces to a half-filled $\jeff=1/2$ single band near the Fermi surface, a configuration which makes 
it prone to the opening of a Mott gap as a result of repulsive interactions.

As the non-interacting Fermi surface of this material is electron-like, the hole-doped regime of high-Tc cuprates is to be compared with the electron-doped one of $\iridate$. 
Several experimental groups performed ARPES measurements on $\iridate$ to investigate the doped compound further.~\cite{kim_science_2014,geneva_prl_2015,brouet_prb_2015,terashima_prb_2017}
Spectral intensity at the Fermi surface exhibits a strong momentum differentiation leading to the appearance of pockets
in the `nodal' region located around $(\pi/2, \pi/2)$,~\cite{geneva_prl_2015,kim_science_2014}
while the ARPES spectra in the `antinodal' region around $(\pi,0)$ are suggestive of a pseudogap.~\citep{geneva_prl_2015}  
Note that the `nodal/antinodal' terminology is inherited from the cuprate context and does not refer to the 
nodes of a superconducting gap - up to now no unambiguous evidence of superconductivity has been established. 

\begin{figure}
\centering
\includegraphics[width=0.45\textwidth]{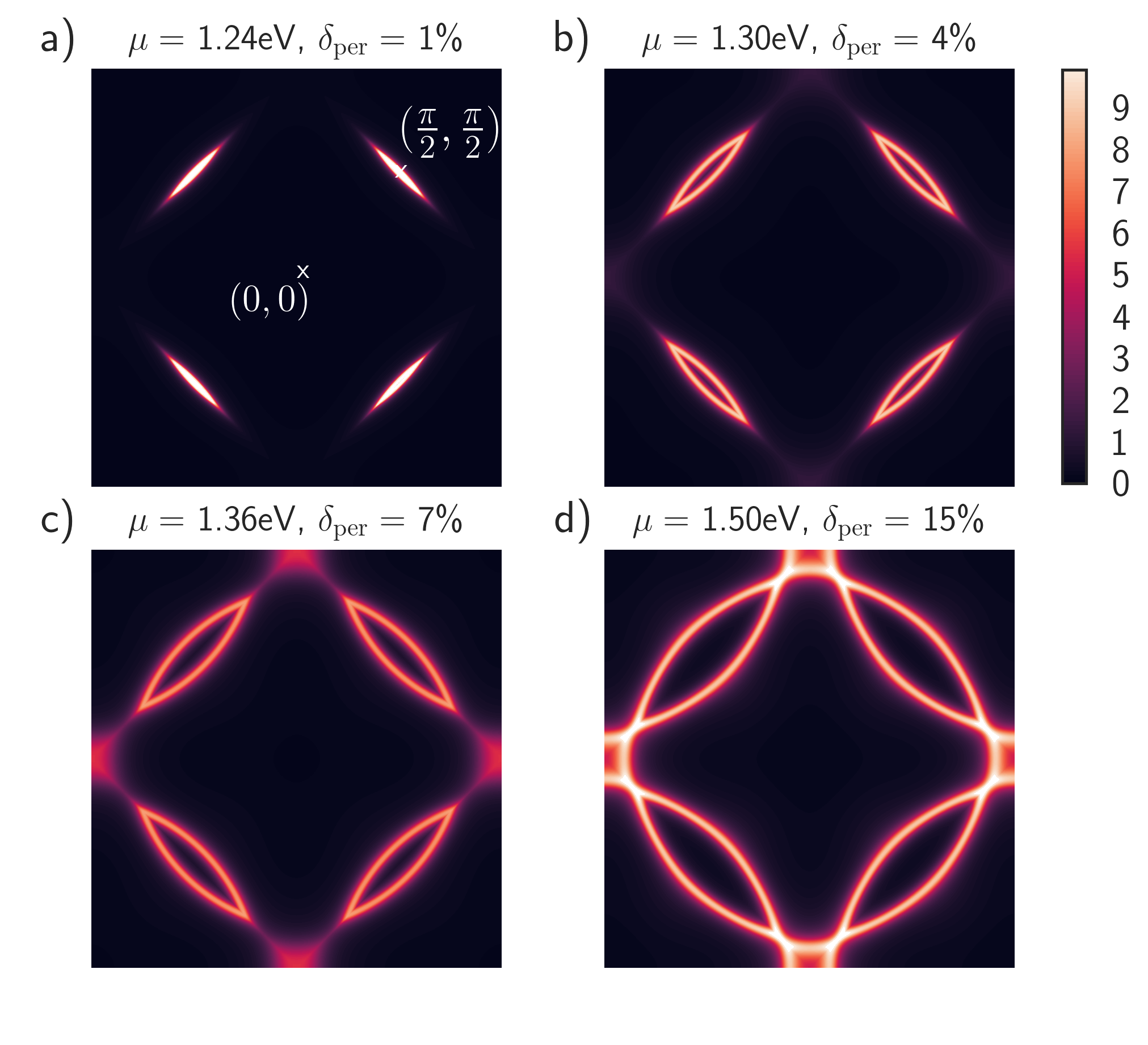}
\caption{Spectral intensity at the Fermi surface illustrating the evolution of the electronic structure upon increasing doping level, 
as described in the text. At low doping, the Fermi surface only consists in pockets near $(\pi/2, \pi/2)$ and a pseudogap is found 
near the antinodes.  Results obtained within CDMFT and a periodized self-energy for $U = 2$ eV at $T=58$~K (see text).
}
\label{Evolution_intensity}
\end{figure}

In this article, we construct a theoretical model of the low-energy electronic structure of doped $\iridate$, 
treating electronic correlation effects in the framework of cellular dynamical mean-field theory (CDMFT).~\cite{cdmft_2001,antoine_rmp_1996}
The Fermi surface spectral intensity maps displayed in Fig.~\ref{Evolution_intensity} summarize key aspects 
of our results. Four successive doping regimes are found: The Mott insulating state (not displayed in Fig.~\ref{Evolution_intensity}) 
evolves into a metal with strong nodal-antinodal differentiation at low doping level. In this regime, the Fermi surface 
consists in pockets around $(\pi/2, \pi/2)$ (a-b), while the antinodal region displays a pseudogap, as shown below. 
Increasing doping further, spectral intensity appears near the antinodes with still a pronounced 
differentiation (c). A full Fermi surface, close to the uniform non-interacting one, is recovered at higher doping (d).

A previous theoretical study~\cite{hampel_prb_2015} addressed the issue of nodal-antinodal differentiation in this material, but 
the opening of the antinodal pseudogap and the associated spectral signatures could not be discussed due to the limitations 
of the slave-boson method used in that work.

This article is organized as follows. In Sec.~\ref{sec:tb_model}, we briefly
summarize the tight-binding (TB) model of the electronic structure established in previous works 
and derive an effective model for the $\jeff = 1/2$ states. In Sec.~\ref{sec:cdmft}, we introduce
correlations in this model and explain how to deal with these within 
the CDMFT scheme. We discuss
results of such CDMFT calculations in Sec.~\ref{sec_four} and particularly
the emergence of the four doping regimes sketched above. In order to restore 
translational symmetry which is broken in CDMFT, we introduce a periodization scheme for the self-energy
in Sec.~\ref{sec:fermiology}. This allows for a calculation of the spectral intensities, which are found to be
in good agreement with the existing ARPES measurements. We also discuss the spectral signatures of 
the antinodal pseudogap. 
In Sec.~\ref{sec:band_structure}, we compute and discuss the the quasiparticle band structure. 
Sec.~\ref{sec:conclusion} is devoted to a discussion and concluding remarks. 

%
%
\section{Tight-binding model} \label{sec:tb_model}

In this section, we describe the electronic band structure of $\iridate$, derive an effective model for
the $\jeff=1/2$ states, and
emphasize that the low-energy states can be described by a single band tight-binding
model defined on a periodic lattice with a single atom per unit cell.

Our starting point is the tight-binding (TB) model introduced in
Refs.~\onlinecite{jin_prb_2009, carter_prb_2013} describing $t_{2g}$
bands in the presence of a spin-orbit coupling.
Because of the rotation of the $\text{IrO}_6$ octahedra around the $c$-axis,
the $\iridate$ unit cell is composed of two inequivalent sites $A$ and $B$.
The tight-binding Hamiltonian is then written as
\begin{equation}
  \mathcal{H}_0 = \sum_{\bk \in \mathrm{RBZ}} \psi^\dagger_{\bk} H_0(\bk) \psi_{\bk},
\end{equation}
where the momentum sum is over the $\sqrt{2}\times\sqrt{2}$ reduced Brillouin
zone and the components of $\psi_\bk$ are the electron annihilation operators
for all 12 orbitals in the unit cell $\{ c_{\bk \tau \alpha \sigma} | \tau=A,B;
\alpha= d_{xy}, d_{yz}, d_{zx}; \sigma=\uparrow, \downarrow\}$. It is
convenient to order the basis according to $(c_{Ad_{xy}\uparrow},
c_{Ad_{yz}\downarrow}, c_{Ad_{zx}\downarrow}, [A \leftrightarrow B])$ followed
by their time-reversed partners $([\uparrow \leftrightarrow \downarrow])$. 
There is no coupling between these two
blocks as
the system is time-reversal invariant and we can thus only consider the first half of the basis, taking into
account that all bands are two-fold degenerate. The remaining $6 \times 6$
tight-binding matrix $H_0$ writes
\begin{equation}
\label{tb_so_hamilt}
  H_0(\bk) = \begin{pmatrix}
  O(\bk) & P(\bk) \\
  P^\dagger(\bk) & O(\bk)
  \end{pmatrix},
\end{equation}
where $P$ describes the hopping part of the Hamiltonian
\begin{equation}
  P(\bk) = e^{-i\frac{k_x + k_y}{2}}\begin{pmatrix}
  -4t_1(\bk) & 0 & 0 \\
  0 & -2t_2(\bk) & 0 \\
  0 & 0 & -2t_3(\bk)
  \end{pmatrix},
\end{equation}
with $t_1(\bk) = t_0\cos \frac{k_x}{2} \cos \frac{k_y}{2}$, $t_2(\bk) = t_0 \cos
\frac{k_x + k_y}{2}$ and $t_3(\bk) = t_0 \cos \frac{k_x - k_y}{2}$.
Here $\bk = (k_x, k_y)$ is expressed in terms of the reciprocal
vectors forming the reduced Brillouin zone.
$O$ describes the on-site part of the Hamiltonian. It includes the spin-orbit
coupling $\lambda \, \textbf{L}_i \cdot \textbf{S}_i$ and reads
\begin{equation}
  O(\bk) = \begin{pmatrix}
  \Delta_t + e_1 (t_1(\bk)/t_0)^2 & \lambda/2 & -i\lambda/2 \\
  \lambda/2 & 0 & -i\lambda/2 \\
  i\lambda/2 & i\lambda/2 & 0
  \end{pmatrix},
\end{equation}
where $\Delta_t$ is an on-site energy difference of the $d_{xy}$ orbital
relative to $d_{yz}$ and $d_{zx}$, and $\lambda$ is the spin-orbit coupling
parameter.  The additional term $e_1 (t_1/t_0)^2$ accounts for the
hybridization between $d_{xy}$ and $d_{x^2-y^2}$.~\cite{jin_prb_2009} In the
following we consider $\Delta_t = 0.15$ eV, $t_0 = 0.35$ eV, $e_1 = -1.5$ eV
and $\lambda=0.57$ eV. It has been shown that these values yield a band
structure in good agreement with LDA+SO calculations.~\cite{jin_prb_2009,
geneva_prl_2015}

\begin{figure}
\centering
\includegraphics[width=0.46\textwidth]{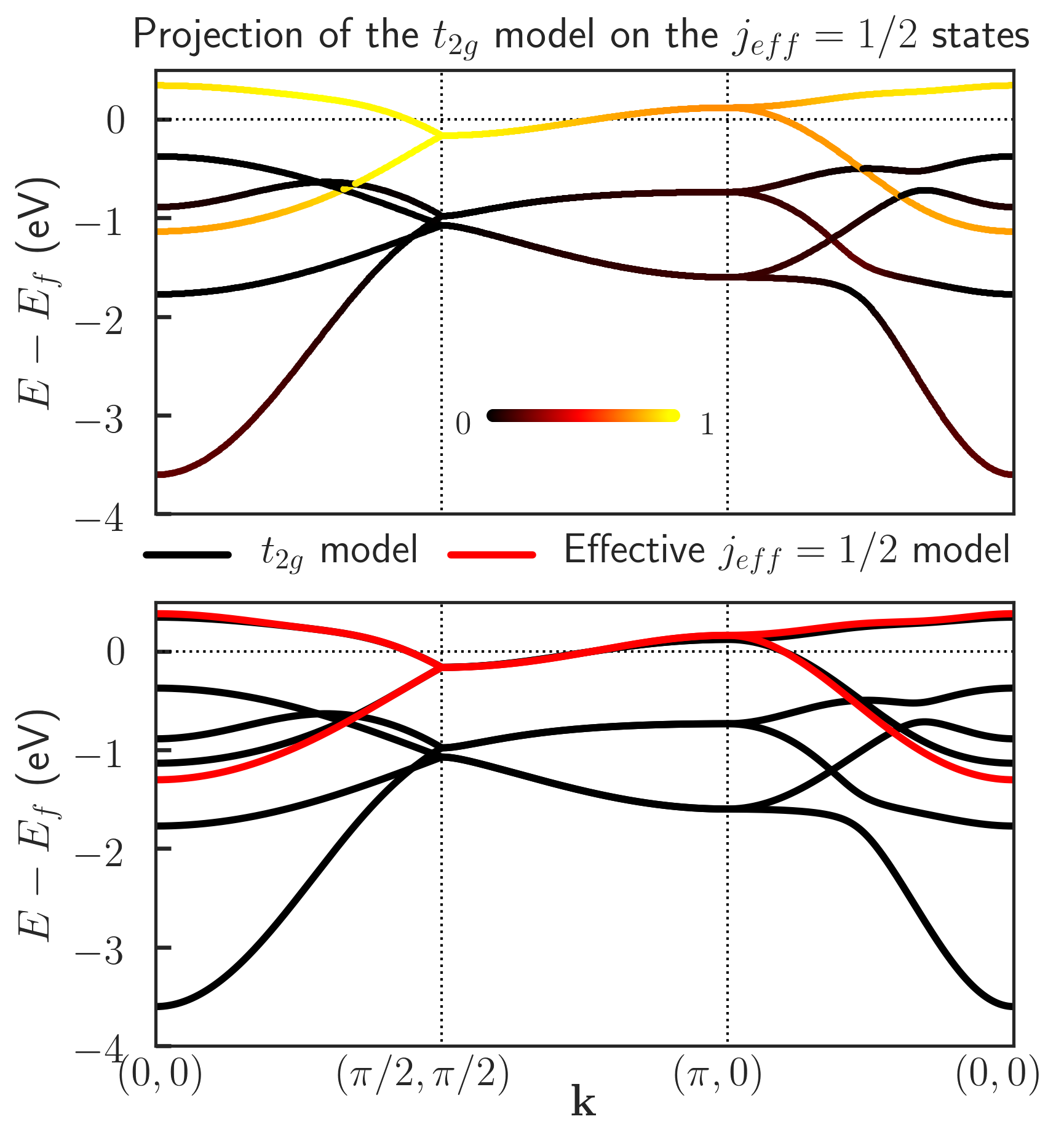}
\caption{
\emph{Upper panel:} Projection of the tight-binding model for the $t_{2g}$
bands (Eq.~\eqref{tb_so_hamilt}) on the $\jeff$ = 1/2 states. The value of the
projection ranges from 0 (black) to 1 (yellow). \emph{Lower panel:} Comparison
between the full $t_{2g}$ model (black lines) and the effective $\jeff = 1/2$
model from Eq.~\eqref{eq:heff} (red lines) with $E=0$. On both panels, bands are
plotted in reciprocal space, along the $(0,0) - (\pi/2, \pi/2) - (\pi,0) -
(0,0)$ path of the full Brillouin zone. $\Delta_t = 0.15$ eV, $t_0 = 0.35$ eV,
$e_1 = -1.5$ eV and $\lambda=0.57$ eV.
}
\label{tb_model}
\end{figure}

We plot in Fig.~\ref{tb_model} (upper panel) the six bands resulting from the
diagonalization of $H_0(\bk)$ along the $(0,0) - (\pi/2, \pi/2) - (\pi,0) - (0,0)$
path of the full Brillouin zone. When the eigenvalues are projected on the $\jeff
= 1/2$ states
\begin{equation}
  \left| \jeff = \frac{1}{2}, \pm \frac{1}{2} \right\rangle = \mp \frac{1}{\sqrt{3}} \left[ |d_{xy}, \pm\rangle \pm (|d_{yz}\rangle,\mp) \pm i|d_{zx},\mp\rangle)\right],
\end{equation}
it can be seen that the low-energy bands essentially have $\jeff=1/2$
character, as highlighted in Refs~\onlinecite{kim_prl_2008, jin_prb_2009,
carter_prb_2013, geneva_prl_2015}.
It is therefore natural to look for an effective reduced
$2\times 2$ Hamiltonian describing these states.

To do so, we rewrite $H_0$ in the basis ($|\frac{1}{2}, \frac{1}{2}\rangle_A$, $|\frac{1}{2}, \frac{1}{2}\rangle_B$, $|\frac{3}{2}, \frac{1}{2}\rangle_A$, $|\frac{3}{2}, \frac{1}{2}\rangle_B$, $|\frac{3}{2}$, $\frac{-3}{2}\rangle_A$, $|\frac{3}{2}, \frac{-3}{2}\rangle_B$):
\begin{equation}
  H_0(\bk) = \begin{pmatrix}
  H_{1/2}(\bk) & M(\bk) \\
  M^\dagger(\bk) & H_{3/2}(\bk)
  \end{pmatrix},
\end{equation}
where the exact expressions of $H_{1/2}$, $M$ and $H_{3/2}$ are given in
Appendix~\ref{app:change_basis}. An effective Hamiltonian
is then obtained by projecting $H_0$ onto the $\jeff=1/2$ subspace:
\begin{equation}
  H_{1/2}^{\text{eff}}(\bk) = H_{1/2}(\bk) + M(\bk)\left[E \times \mathbf{1}_{4\times4} - H_{3/2}(\bk)\right]^{-1}M^\dagger(\bk),
\label{eq:heff}
\end{equation}
where $\mathbf{1}_{4\times4}$ is the $4\times 4$ unit matrix and $E$ an energy
scale that is adjusted in order to best match the original band structure.

While it is difficult to have a compact expression for this reduced
Hamiltonian, one can easily diagonalize $H_{1/2}^{\text{eff}}$ numerically for
every $\bk$ point of interest.
This is shown (red lines) in Fig.~\ref{tb_model} (lower panel) together with
the complete $t_{2g}$ band structure (black lines). The effective model appears
to be in excellent agreement with the two low-energy bands exhibiting a $\jeff = 1/2$
character.

\begin{figure}
\centering
\includegraphics[width=0.46\textwidth]{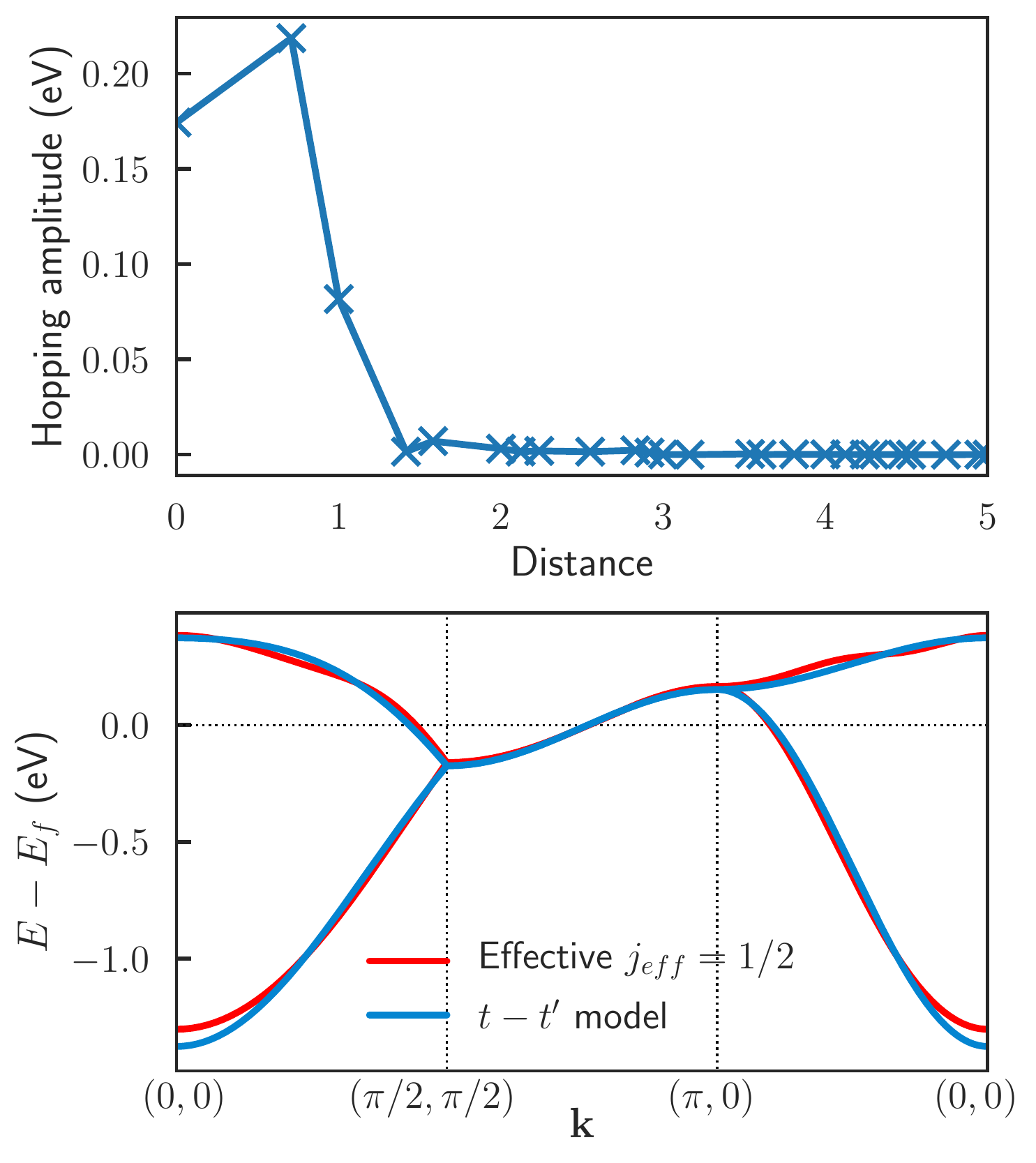}
\caption{\emph{Upper panel:} Real-space hopping amplitudes of effective $\jeff=1/2$ Hamiltonian the effective with 
respect to the distance (the inter-site distance is normalized to 1). 
\emph{Lower panel:} Comparison between the effective $\jeff = 1/2$ bands (red) with $E=0$ and the folded 
dispersion obtained by keeping only the nearest and next-nearest neighbor hopping terms 
($t = -0.219$eV, $t' = -0.082$eV respectively). Bands are plotted in reciprocal space, along the 
$(0,0) - (\pi/2, \pi/2) - (\pi,0) - (0,0)$ path of the full Brillouin zone. 
$\Delta_t = 0.15$ eV, $t_0 = 0.35$ eV, $e_1 = -1.5$ eV and $\lambda=0.57$ eV.}
\label{t_tp_model}
\end{figure}

As mentioned earlier, the $\iridate$ crystal has a two-atom unit cell
and we expressed the tight-binding models above in
the reduced Brillouin zone in order
to make contact with experiments. Let us however emphasize that all sites are actually equivalent
from a purely
electronic point of view in these models.  An
inspection of the band structure in the reduced Brillouin zone indeed reveals
that it results from the folding of half as many bands defined over the
full Brillouin zone. This can be seen e.g. from the degeneracy of the bands
along the $(\pi/2,\pi/2)$ - $(\pi,0)$ path of the full
Brillouin zone. As a result, the effective model in Eq.~\eqref{eq:heff}
can be written as a simple tight-binding model on a square lattice
\begin{equation}
  \mathcal{H}^\mathrm{eff}_{1/2} = \sum_{ij} t_{ij} c^\dagger_i c_j,
\end{equation}
where the hopping amplitudes $t_{ij}$ are shown as a function of the
distance $|i-j|$ in Fig.~\ref{t_tp_model}
(upper panel). We see
that one obtains a good approximation of the band structure by only keeping the nearest and
next-nearest neighbor hopping terms (for an almost perfect agreement it is necessary to
keep 8 hopping parameters). This yields the simple energy dispersion
\begin{equation}
  \epsilon(\bk) = \epsilon_0 + 2t(\cos k_x + \cos k_y) + 4t'\cos k_x \cos k_y,
\end{equation}
where $\epsilon_0 = -0.174$ eV, $t = -0.219$ eV, $t' = -0.082$ eV  
and $\bk =(k_x, k_y)$ is now expressed in the basis of the \emph{full} Brillouin zone.
The folding of this band in the reduced Brillouin zone is shown together with
the effective $\jeff=1/2$ band structure previously derived in
Fig.~\ref{t_tp_model} (lower panel). Let us mention that a similar
tight-binding model was derived in Ref.~\onlinecite{watanabe_prl_2010} with the
difference that the $d_{x^2 - y^2}$ admixture was not included in their work.

In the following we use the effective Hamiltonian $\mathcal{H}^\mathrm{eff}_{1/2}$ to
describe the low-energy excitations of the system.

%
%
\section{Introducing correlations} \label{sec:cdmft}

We model the effect of electronic correlations in $\iridate$ with a
Hubbard Hamiltonian that introduces an energy cost for having two electrons on the
same Ir atom
\begin{equation}
  \mathcal{H} = \mathcal{H}^\mathrm{eff}_{1/2} + U \sum_{i, \tau} n_{i \tau \uparrow} n_{i \tau \downarrow},
\label{eq:hubbard}
\end{equation}
where $n_{i \tau \sigma}$ is the occupation number on the $j_z = \sigma$
orbital of the Ir atom $\tau=A,B$ in the unit cell $i$. In the following we use
$U = 2$~eV~\cite{jin_prb_2009,geneva_prl_2015} and temperature $T = 1/\beta =
1/200$~eV $\simeq 58$~K.
This model is studied using cellular dynamical mean-field
theory~\cite{cdmft_2001,antoine_rmp_1996}: The original lattice
Hamiltonian~\eqref{eq:hubbard} is mapped on a two-site auxiliary cluster model
embedded in a self-consistent medium.
The self-energy of the cluster model $\Sigma_{\tau, \tau'}$ is used to
construct an approximation of the lattice self-energy where only intra unit
cell components are non-vanishing, i.e.  $\Sigma^\mathrm{latt}_{i\tau,i\tau'} =
\Sigma_{\tau, \tau'}$. Note that the orbitals at sites $A$ and $B$ are
electronically equivalent and therefore $\Sigma_{AA} = \Sigma_{BB}$ and
$\Sigma_{AB} = \Sigma_{BA}$. We then have the following expression for the lattice
Green's function
\begin{equation}
  \hat{G}^\mathrm{latt}(i\omega_n,\bk) = \Big\{ (i\omega_n + \mu) \mathbf{1} - H^\mathrm{eff}_{1/2}(\bk)
    - \hat{\Sigma}(i\omega_n) \Big\}^{-1},
\label{eq:glattice}
\end{equation}
where $\bk$ is defined in the reduced Brillouin zone and both
$\hat{G}^\mathrm{latt}$ and $\hat{\Sigma}$ are $2\times 2$ matrices
associated with the two Ir atoms in the unit cell.
The CDMFT self-consistency imposes that the cluster Green's function
$\hat{G}$ be the same as the unit cell Green's function of the lattice:
\begin{equation}
  \hat{G}(i\omega_n) = \sum_{\bk \in \mathrm{RBZ}} \Big\{ (i\omega_n + \mu) \mathbf{1} - H^\mathrm{eff}_{1/2}(\bk)
  - \hat{\Sigma}(i\omega_n) \Big\}^{-1}.
\label{eq:cdmft}
\end{equation}
We use a continuous-time quantum Monte Carlo (CT-HYB)~\cite{seth_cthyb_2016,
werner_prl_2006, werner_prb_2006, gull_rmp_2011} impurity solver to find the
solution of the two-site cluster model and the self-consistent
equation~\eqref{eq:cdmft} is solved iteratively.~\cite{antoine_rmp_1996}
More details are given in Appendix~\ref{app:cdmft}. 
Codes necessary for the numerical calculations were 
developed using the TRIQS~\cite{triqs_2015} library. 

%
%
\section{The four doping regimes} \label{sec_four}

We first investigate the cluster quantities $\hat{G}$ and $\hat{\Sigma}$
obtained by solving the CDMFT equations. These quantities can be expressed
in the basis $\{ | \frac{1}{2}, \frac{1}{2} \rangle_A, |\frac{1}{2}, \frac{1}{2} \rangle_B \}$
of the $\jeff=1/2$ orbitals on sites $A$ and $B$. However, because $A$ and $B$ are electronically
equivalent, it is convenient to work in the basis $\{ |+\rangle, |-\rangle \}$
of even and odd combinations of the $\jeff=1/2$ orbitals, defined by
\begin{equation} \label{eq_pm}
  \Big| \pm \Big\rangle = \frac{1}{\sqrt{2}} \left( \left|\frac{1}{2}, \frac{1}{2} \right\rangle_A \pm \left|\frac{1}{2}, \frac{1}{2} \right\rangle_B \right).
\end{equation}
In this basis, both $\hat{G}$ and $\hat{\Sigma}$ are diagonal (see Appendix~\ref{app:cdmft})
\begin{equation}
  \hat{G} = \begin{pmatrix} G_+ & 0 \\ 0 & G_- \end{pmatrix} \qquad
  \hat{\Sigma} = \begin{pmatrix} \Sigma_+ & 0 \\ 0 & \Sigma_- \end{pmatrix}.
\end{equation}
As we will discuss later, $G_\pm$ and $\Sigma_\pm$ have a direct physical
interpretation. The physics close to the node $(\pi/2,\pi/2)$ is indeed
essentially controlled by $G_-$ and $\Sigma_-$ while the physics at the
antinode $(\pi,0)$ is controlled by $G_+$ and $\Sigma_+$. 
The reason for this, anticipating on Sec.~\ref{sec:band_structure} and Fig.~\ref{bands_article}, is that 
the nodal Fermi-surface pocket at $(\pi/2,\pi/2)$ is associated with the upper band (which has an antibonding/odd character)
while the nodal states are associated with the lower bonding band with even character.
The analysis of these quantities will reveal the existence of four distinct regimes upon doping: a
Mott insulator phase, a pseudogap regime, a differentiation region and finally a
uniform Fermi liquid state.

\begin{figure}
\centering
\includegraphics[width=0.45\textwidth]{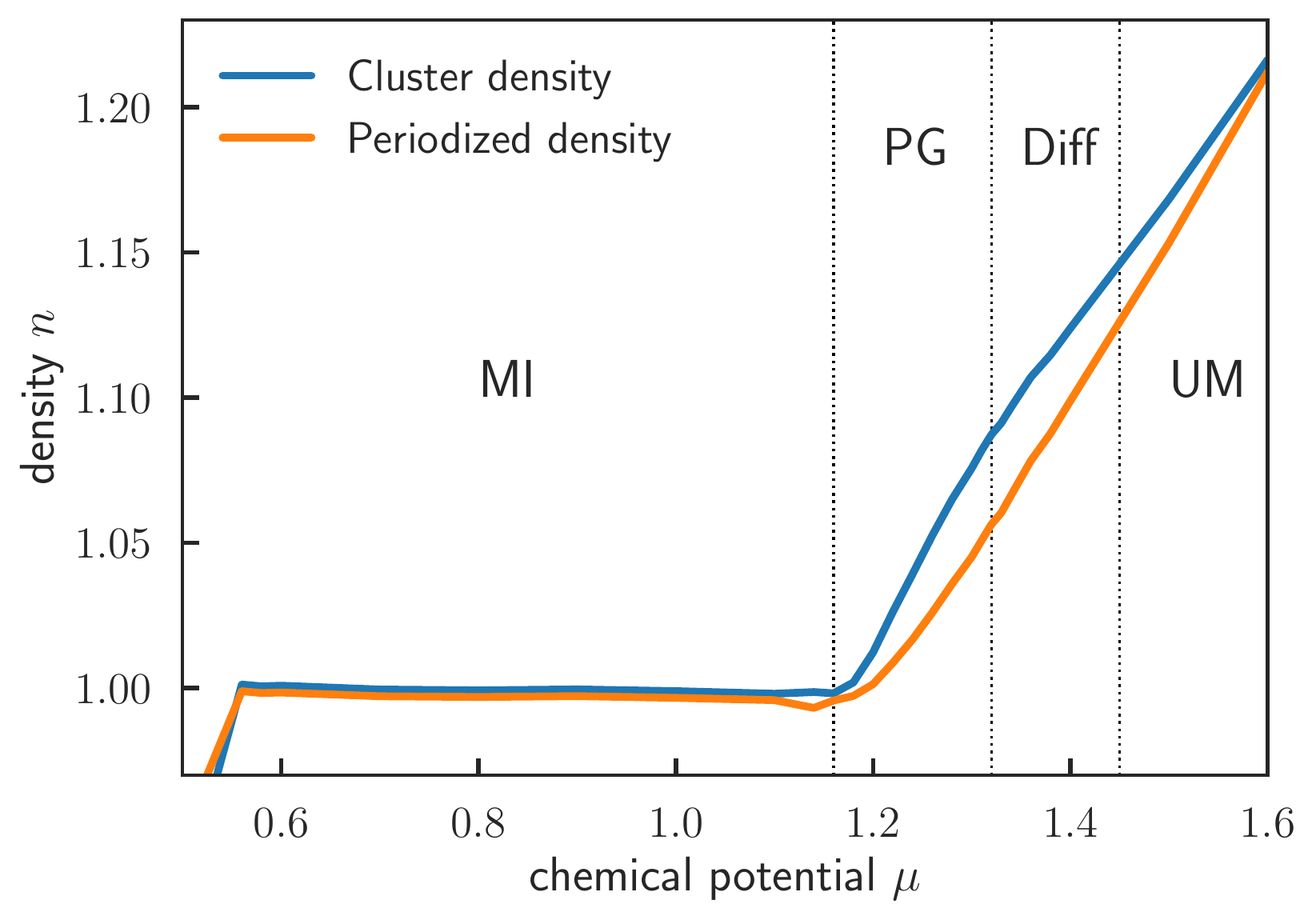}
\caption{Density of states as a function of the chemical potential $\mu$. The
cluster density is plotted in blue while the periodized one is plotted in
orange. Dotted lines separate the four doping regimes: below $\mu=1.16$eV is the Mott
insulating phase (MI), between $\mu=1.16$eV and $\mu=1.32$eV is the pseudogap
regime (PG), between $\mu=1.32$eV and $\mu=1.45$eV is the differentiation region
(Diff), and above $\mu=1.45$eV is the uniform metal (UM). Results obtained with a 2-site CDMFT calculation 
for $U=2$ eV, $T=58$ K.}
\label{Densities_beta200_article}
\end{figure}

The electronic density $n$ is shown as a function of the chemical potential
$\mu$ in Fig.~\ref{Densities_beta200_article} (blue curve). It displays a clear
plateau at $n=1$ for $\mu$ between $0.56$eV and $1.16$eV, confirming that the system
is a Mott insulator at half-filling.~\cite{kim_prl_2008,martins_prl_2011,zhang_prl_2013} The
width of the plateau $\simeq 0.6$eV is consistent with the recent experiment
of Ref.~\onlinecite{brouet_prb_2015}.

\begin{figure}
\centering
\includegraphics[width=0.45\textwidth]{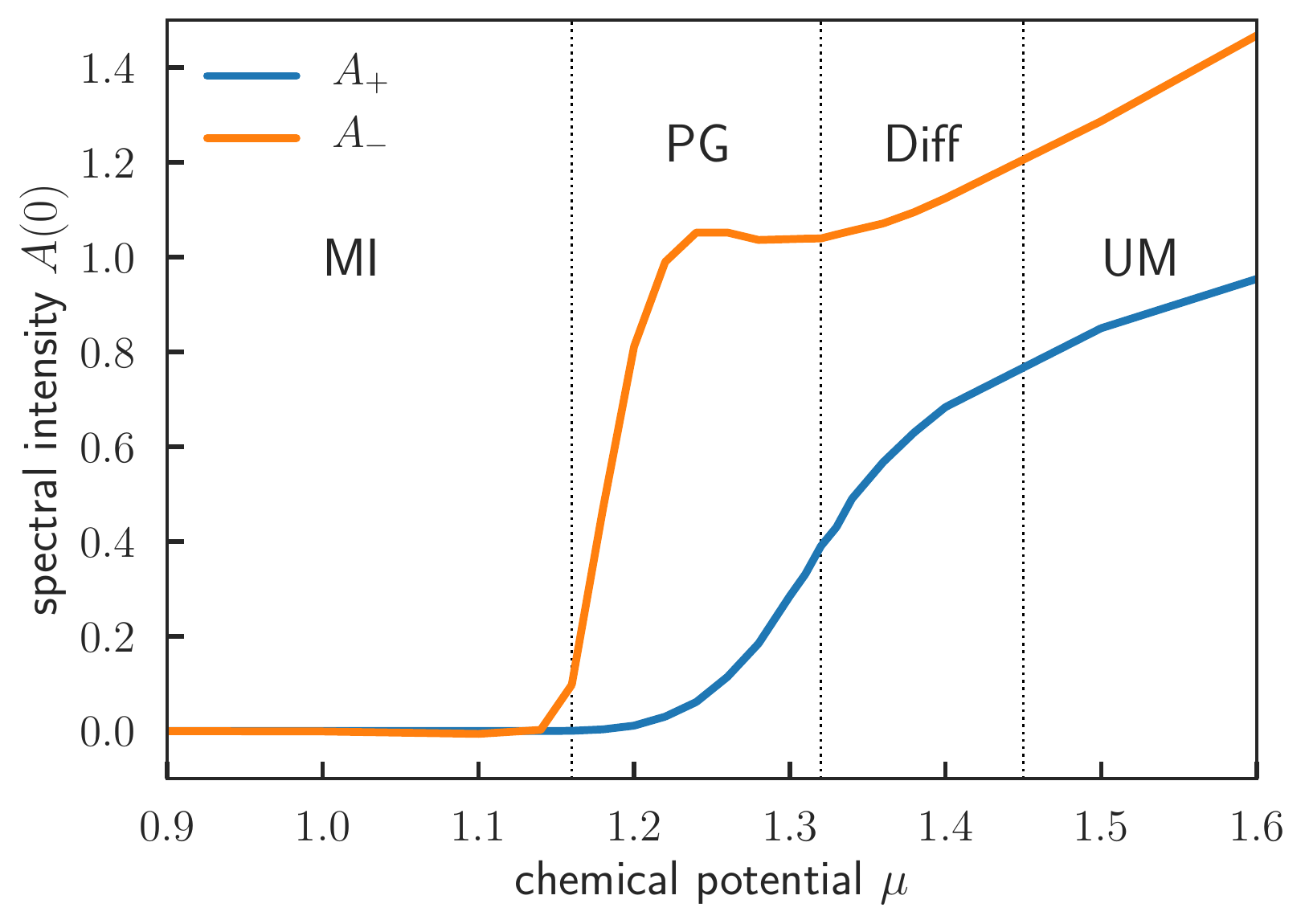}
\caption{Spectral intensity at the Fermi level $A(0)$ as a function of the
chemical potential $\mu$. The even (odd) contribution is plotted in blue
(orange). Dotted lines separate the four doping regimes (see Fig.~\ref{Densities_beta200_article}).
Results obtained with a 2-site CDMFT calculation for $U=2$ eV, $T=58$ K.}
\label{Extrapolation_Spectral_beta200}
\end{figure}

\begin{figure}
\centering
\includegraphics[width=0.45\textwidth]{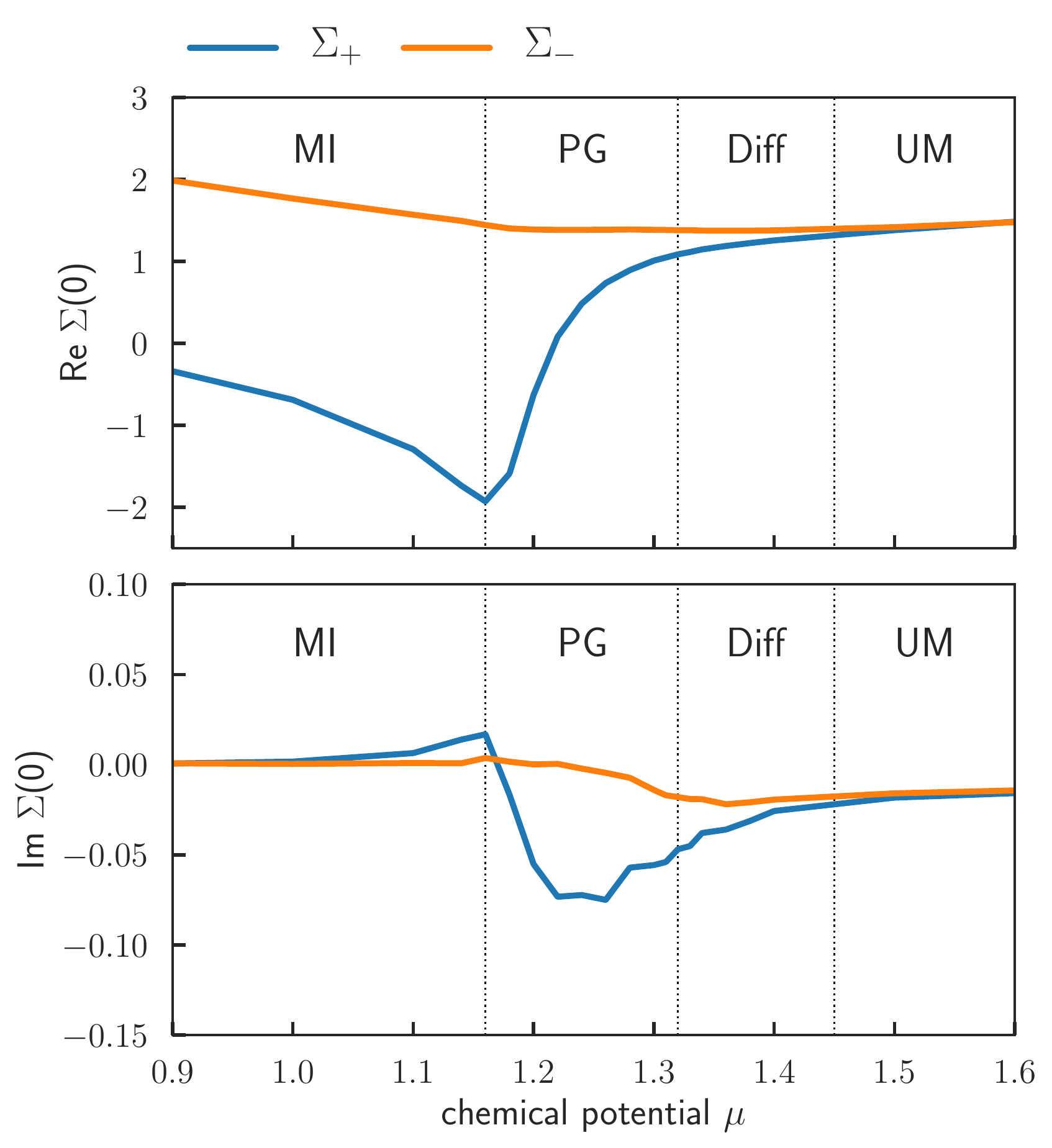}
\caption{Extrapolation at zero frequency of the real part (\emph{upper panel}) and of the imaginary part (\emph{lower panel}) 
of the self-energy $\Sigma(i\omega_n)$ as a function of the chemical potential $\mu$. 
The even (odd) contribution is plotted in blue (orange). 
Dotted lines separate the four doping regimes (see Fig.~\ref{Densities_beta200_article}).
Results obtained with a 2-site CDMFT calculation for $U=2$ eV, $T=58$ K.}
\label{Extrapolation_Sigma_beta200_article}
\end{figure}

In Fig.~\ref{Extrapolation_Spectral_beta200} and
~\ref{Extrapolation_Sigma_beta200_article} are displayed the spectral
intensities $A_\pm(\omega=0)$ at the Fermi level as well as the zero-frequency
self-energies $\Sigma_\pm(\omega=0)$ as a function of the chemical potential
$\mu$. These quantities are obtained by extrapolating to zero Matsubara
frequencies results obtained by Monte Carlo:
\begin{align}
  A_\pm(0) &= -\frac{1}{\pi} \lim_{i\omega_n \rightarrow 0} \text{Im} G_\pm(i\omega_n),
  \\
  \Sigma_\pm(0) &= \lim_{i\omega_n \rightarrow 0} \Sigma_\pm(i\omega_n).
\end{align}
For completeness, we have included plots of the Matsubara frequency Green's functions
and self-energies for several chemical potentials in Appendix~\ref{app:matsubara}.

These results allow to identify four distinct doping regimes.  For chemical
potentials smaller than $\mu=1.16$eV, the system is in a Mott insulating regime
and both the even ($+$) and odd ($-$) components of the spectral intensity at
the Fermi level are zero, $A_\pm(0) = 0$ (also both Matsubara Green's functions
$G_\pm(i\omega_n)$ have clear insulating character, see
Appendix~\ref{app:matsubara}).  This is compatible with the location of the
Mott plateau in Fig.~\ref{Densities_beta200_article}. Correlation effects
are especially visible in the very different values of the real parts of the
self-energies while both
imaginary parts vanish. As a result, the effective low-energy band structure is
split by the real parts of the self-energy in Eq.~\eqref{eq:cdmft} and no
excitations exist at $\omega=0$. More precisely, the quasiparticle equation
\begin{equation}
 \det \Big\{ (\omega + \mu) \mathbf{1} - H^\mathrm{eff}_{1/2}(\bk)
  - \hat{\Sigma}(\omega) \Big\} = 0
\end{equation}
has no solutions at $\omega=0$ for all values of $\bk$.

When $\mu$ lies between 1.16eV and 1.32eV, we enter a \emph{pseudogap}
regime. The even component of the Green's function, that provides a
coarse-grained picture of the physics close to the antinode $\bk=(\pi,0)$,
maintains its insulating character ($A_+(0) =0$) while the odd component,
describing the nodal region close to $\bk=(\pi/2,\pi/2)$, becomes metallic
($A_-(0) \neq 0$). This describes a metal that only has coherent quasiparticles
close to the node. Antinodal particles are suppressed by lifetime effects, as
can be seen from the more negative imaginary part of the even self-energy
$\mathrm{Im} \Sigma_+(0)$ reaching -0.1eV in
Fig.~\ref{Extrapolation_Sigma_beta200_article} while $\mathrm{Im} \Sigma_-(0)$
remains very small.  We show below that  the spectral
function exhibits a pseudogap at $\bk = (\pi, 0)$ in this region.  This regime is very reminiscent
of the pseudogap region of cuprate superconductors.

As the electron doping is further increased, for $1.32 \le \mu \le 1.45$eV,
spectral weight starts appearing in $A_+(0)$, an indication that quasiparticles
start forming at the antinode as well. However, there are still visible differences
between the even and odd components of the self-energies (see also Appendix~\ref{app:matsubara}).
The regime is therefore characterized by a visible $\bk$-space differentiation
where lifetime effects are stronger at the antinode than at the node ($\mathrm{Im}\Sigma_+(0) <
\mathrm{Im}\Sigma_-(0)$) but do not completely destroy quasiparticles.

Eventually, for $\mu$ above 1.45eV, a uniform metallic regime settles where both
self-energies are identical and $\bk$-space differentiation has disappeared.
This regime would be well described by a single-site DMFT calculation.

It should be emphasized that boundaries delimiting these different regimes 
correspond to crossovers and hence are defined here in a qualitative manner.

\begin{figure}
\centering
\includegraphics[width=0.45\textwidth]{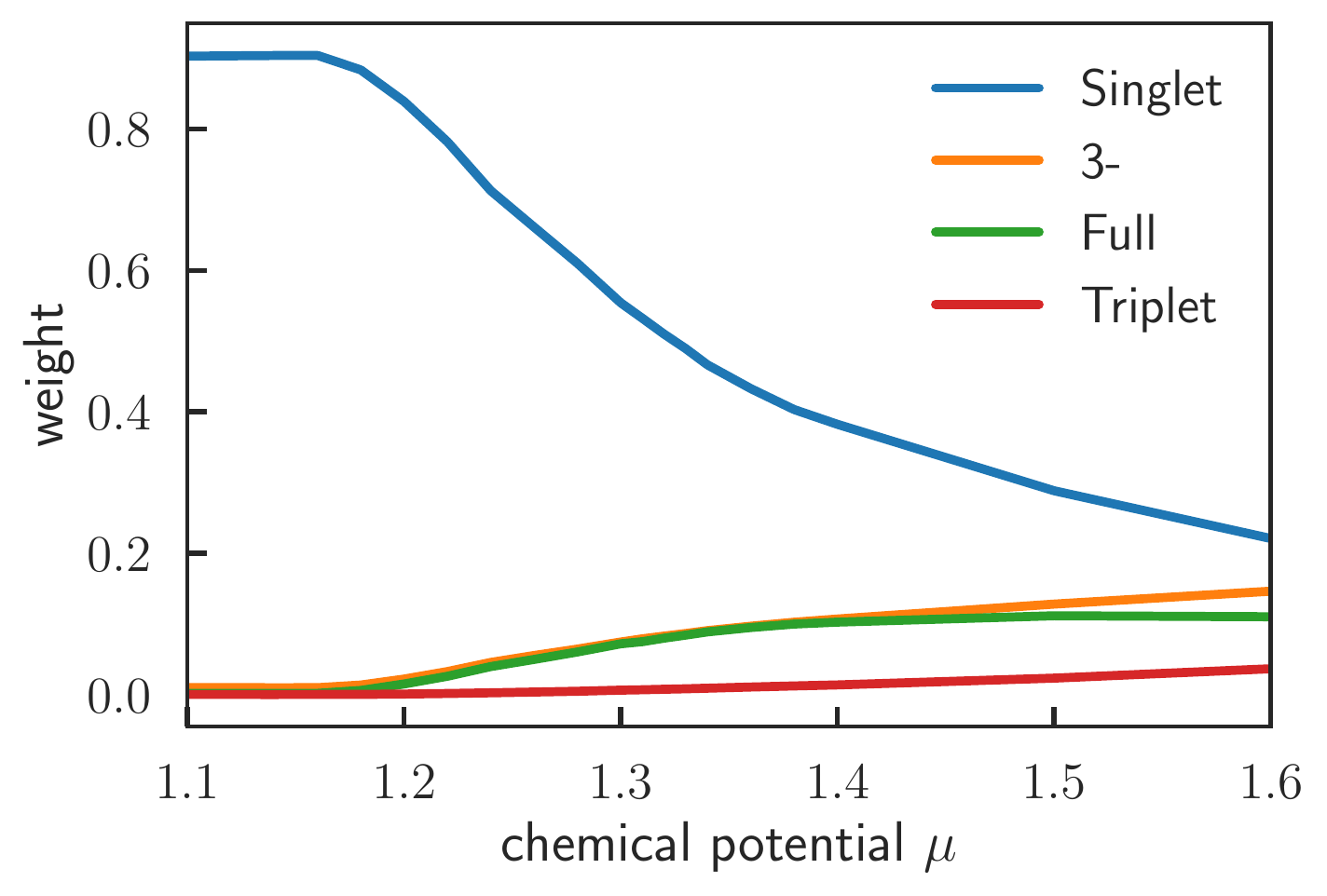}
\caption{Statistical weights of states dominating the Monte Carlo sampling on 
the dimer cluster of the CDMFT approach, as a function of the chemical  potential $\mu$. 
The dominant  state is found to be the 2-electron inter-site singlet (blue).  
As doping level is increased, the three-electrons odd parity state catches up, denoted here as $3-$ (orange) - 
as well as the fully occupied state (green), while the 2-electron triplet state (red) has a smaller weight.
Other contributions are negligible.}
\label{histos_articles}
\end{figure}

The physical mechanism responsible for the formation of the pseudogap and 
the strong nodal-antinodal dichotomy observed at low doping can be revealed 
by studying the many-body states associated with the 2-site cluster.
Calculating these states' histogram, we identify those that contribute
most to the stochastic sampling within the CT-HYB quantum impurity solver. This
is shown in Fig.~\ref{histos_articles}, from which it is clear that
the system is dominated by the intra-dimer singlet state at low doping levels. 
This is a strong indication that physics in this regime is governed by the formation of 
short-range antiferromagnetic correlations between neighboring sites. 

%
%
\section{Fermi surface and pseudogap}\label{sec:fermiology}

We now turn to the study of the fermiology of the system. Within CDMFT, the
lattice Green's function given by Eq.~\eqref{eq:glattice} breaks translational
symmetry,~\cite{cdmft_2001} hence making a direct comparison 
to momentum-resolved ARPES experiments difficult. The reason for the symmetry breaking is  the
lattice self-energy in CDMFT only having components inside a unit cell but not between different
unit cells. A natural way to restore the translational symmetry is to
\emph{periodize} the self-energy by propagating the intersite contribution
$\Sigma_{AB}$ over all links on the lattice.  However an artifact of this
periodization scheme is that it prevents the formation of a Mott insulator and
gives a wrong description of the low-doping physics (see
Appendix~\ref{app:periodization} for more details). We therefore design a different
periodization that yields much more physical results and preserves the existence of the
Mott insulator. In this scheme, the lattice self-energy is given by
\begin{equation}
  \tilde{\Sigma}^\mathrm{latt}(i\omega_n, \bk) = \begin{pmatrix}
  \Sigma_{AA} & \Sigma_{AB} \times e^{-i\frac{k_1+k_2}{2}} \\
  \Sigma_{AB} \times e^{i\frac{k_1+k_2}{2}} & \Sigma_{AA}
  \end{pmatrix},
\end{equation}
where $\bk = (k_1,k_2)$ is expressed in the reduced Brillouin zone.
With this self-energy, we then define a periodized 
lattice Green's function $\tilde{G}^\mathrm{latt}$ according to
\begin{equation}
  \tilde{G}^\mathrm{latt}(i\omega_n,\bk) = \Big\{ i\omega_n + \mu - H^\mathrm{eff}_{1/2}(\bk)
   - \tilde{\Sigma}^\mathrm{latt}(i\omega_n, \bk) \Big\}^{-1}.
\end{equation}
This Green's function preserves all the symmetries of the lattice
and will be the basis of our analysis below.

As a consistency check we first compute in Fig.~\ref{Densities_beta200_article} the electronic density $n$ as a
function of $\mu$ obtained from $\tilde{G}$ (orange curve).  Comparing it to the cluster density
(blue curve) discussed in Sec.~\ref{sec_four}, we see that plateaus at
$n=1$ match well, confirming the existence of a Mott insulator within our
periodization scheme. However, the periodized density generally has a slightly
lower values compared to the cluster density for a given chemical
potential. In the following, we discuss our results for specific values of
$\mu$ and thus indicate two corresponding values of the electron doping: the
cluster and the periodized one (resp. $\delta_{\text{cluster}}$ and
$\delta_{\text{per}}$).

\begin{figure}
\centering
\includegraphics[width=0.45\textwidth]{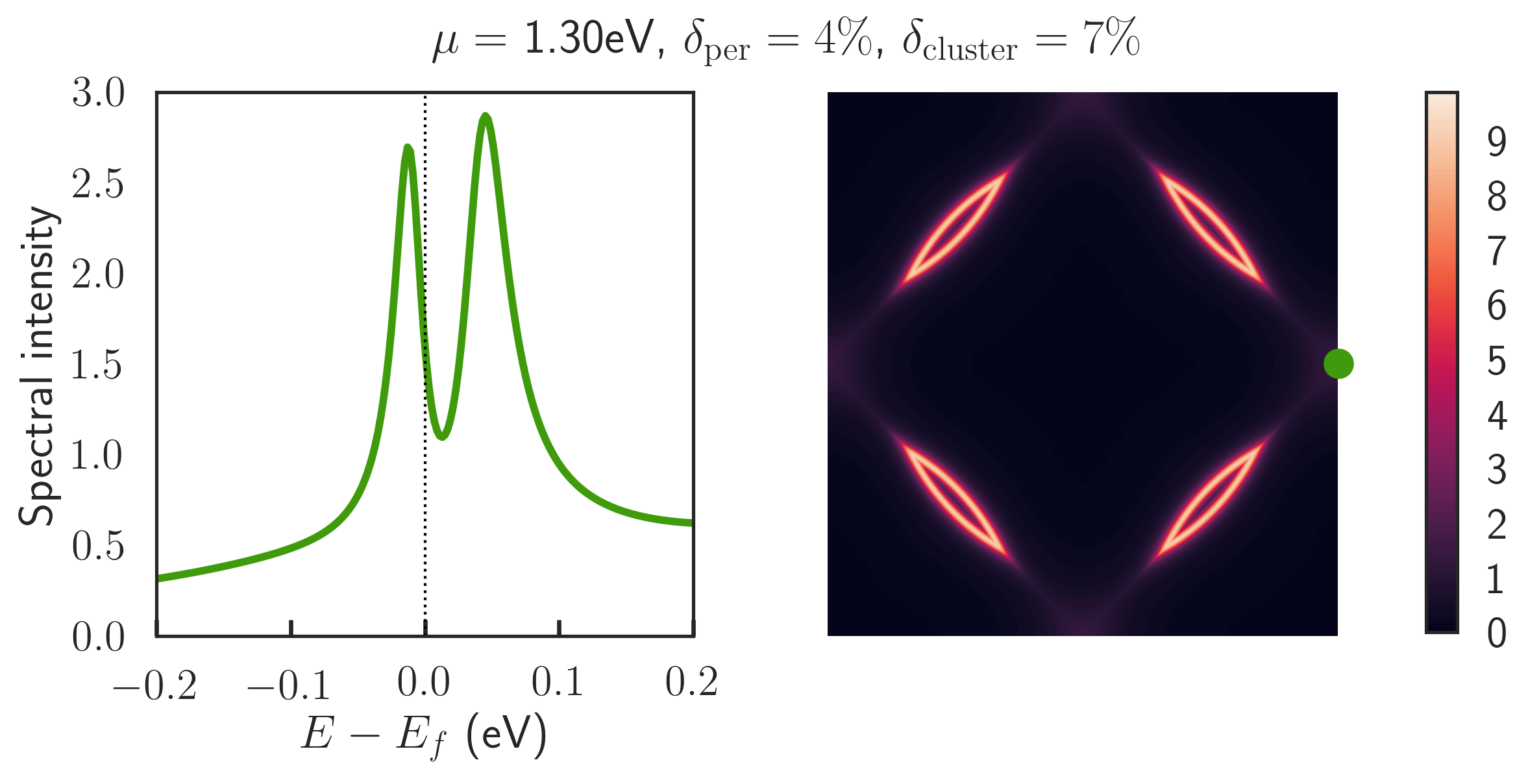}
\caption{\emph{Left panel:} Spectral intensity $\mathrm{Tr} A(\omega, \bk_\mathrm{AN})$ (Energy Distribution Curve - EDC) 
at the antinode $\bk_\mathrm{AN} = (\pi,0)$ for $\mu = 1.3$ eV. \emph{Right panel:} Spectral intensity at the Fermi surface
with a periodized self-energy for the same $\mu$.
$U=2$ eV, $T=58$ K.}
\label{Pseudogap}
\end{figure}

We plot in Fig.~\ref{Evolution_intensity} the spectral intensity at the Fermi
surface for four values of the chemical potential. At small doping levels, for $\mu
\le 1.30eV$, nodal pockets with coherent quasiparticles develop while the
antinodal intensity is completely suppressed. For these values of $\mu$, we are
in the pseudogap regime discussed above. A closer inspection of the
spectral function at $\bk = (\pi,0)$ for $\mu = 1.30eV$ indeed confirms the
presence of a clear pseudogap: Fig.~\ref{Pseudogap} shows the leading edge of the 
spectrum being shifted away from zero energy. As discussed above,  we attribute
its formation to short-range antiferromagnetic correlations (manifested here as 
the dominance of inter-site singlet dimer formation in our cluster, as revealed by the 
histogram of states). 

\begin{figure}
\centering
\includegraphics[width=0.45\textwidth]{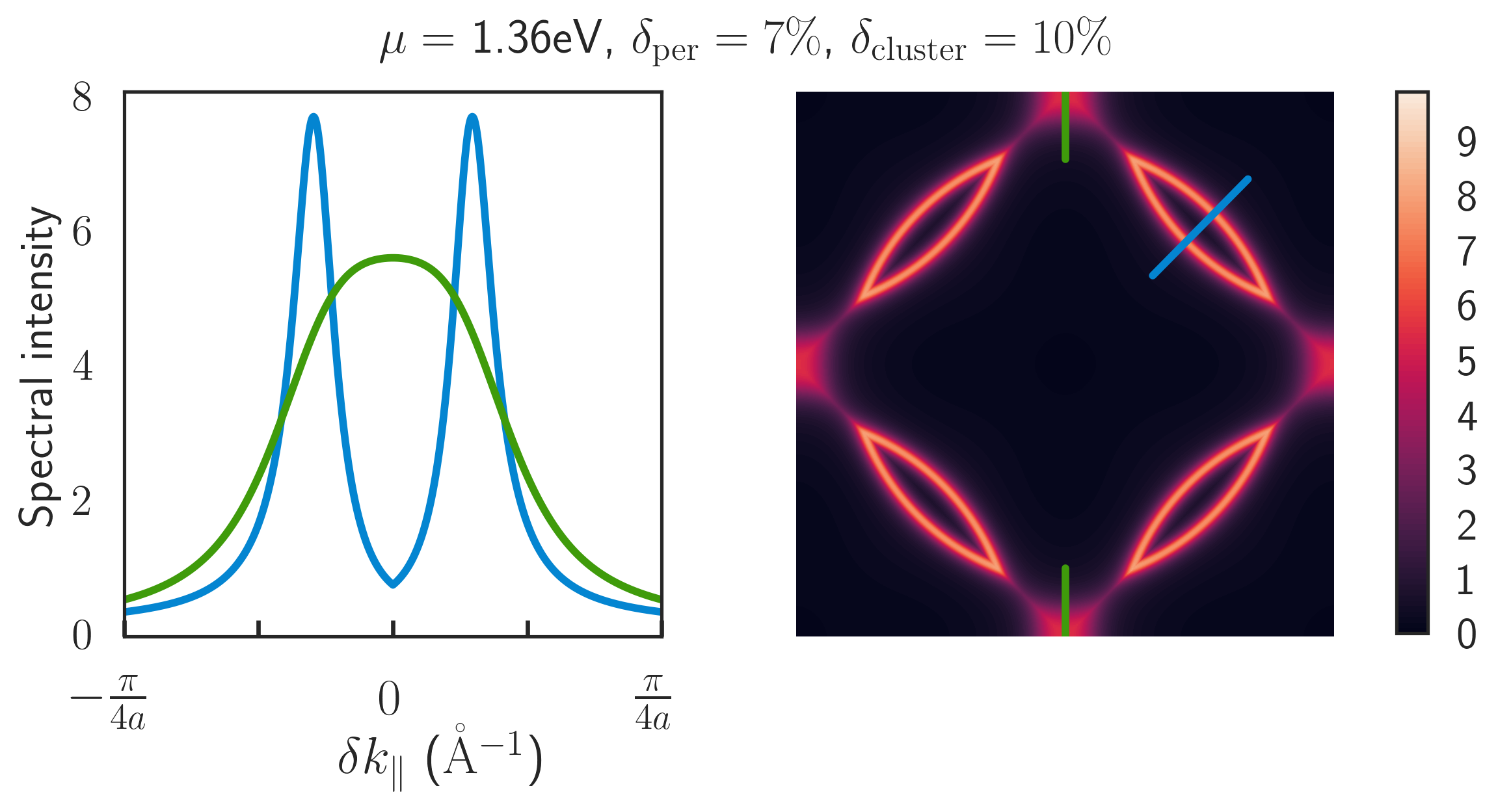}
\caption{\emph{Left panel:} Spectral intensity at the Fermi surface 
$\mathrm{Tr} A(\omega=0, \bk)$ (Momentum Distribution Curve - MDC)
for $\mu=1.36$ eV taken along the nodal (blue)
and the antinodal (green) directions. Corresponding cuts are shown with the
same color code on the right panel.
\emph{Right panel:} Spectral intensity at the Fermi surface
with a periodized self-energy for the same $\mu$.
$U=2$ eV, $T=58$ K.}
\label{Intensity_cut}
\end{figure}

As the electron doping is increased, the $(\pi/2,\pi/2)$
pockets grow and spectral intensity starts to appear around $(\pi,0)$, see panel
(c) of Fig.~\ref{Evolution_intensity}, leading to an extension of the Fermi surface over the
Brillouin zone. Quasiparticles are however far more incoherent and broader
at the antinode, as can be seen from momentum cuts across the node or the
antinode (Fig.~\ref{Intensity_cut}). While sharp coherent quasiparticles are
found at the node, those at the antinode display a lower spectral intensity that
is broadened over a greater region of $\bk$-space. This corresponds to the
momentum-differentiation regime introduced above.

At larger doping, the self-energy becomes finally uniform and the resulting
Fermi surface displays coherent quasiparticles both at the node and the
antinode, as shown in the panel (d) of Fig.~\ref{Evolution_intensity}.

\begin{figure*}
\centering
\includegraphics[width=\textwidth]{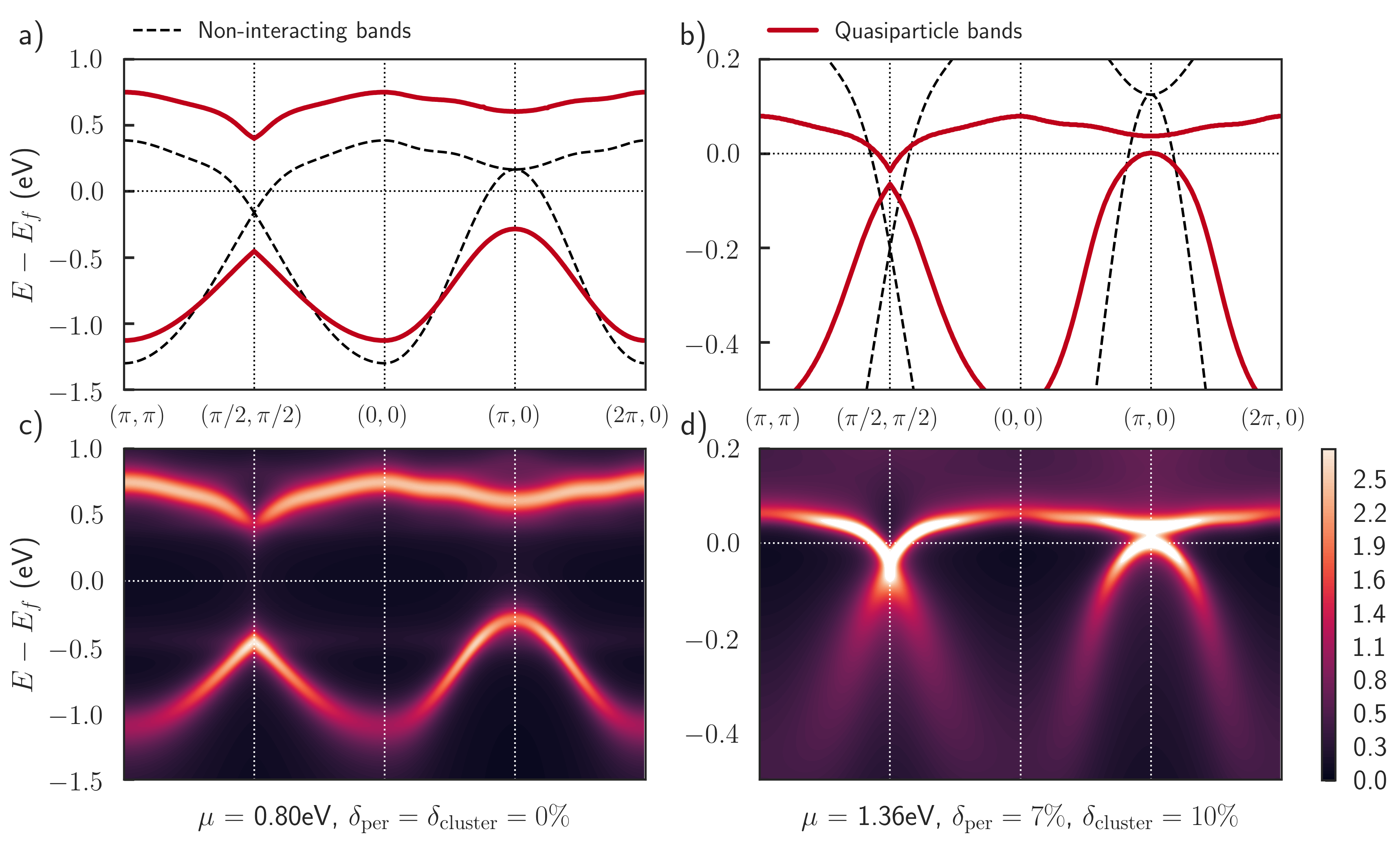}
\caption{Band dispersion of the insulating (a-c), $\mu = 0.8$eV, and doped
(b - d), $\mu = 1.36$eV, $\iridate$. Periodized
self-energies are analytically continued from Pad\'e approximants. \emph{Upper
panels:} Comparison between the non-interacting bands obtained from the TB + SO
Hamiltonian $H_{1/2}^{\text{eff}}$ (dashed lines) and the quasiparticle bands.
\emph{Lower panels:} Spectral intensities, generalizing
Fig.~\ref{Evolution_intensity} away from the Fermi surface. All panels follow
the $(\pi,\pi)$ - $(\pi/2,\pi/2)$ - $(0,0)$ - $(\pi,0)$ - $(2\pi,0)$ path in
the full Brillouin zone. $U=2$ eV, $T=58$ K.}
\label{bands_article}
\end{figure*}

%
%
\section{Electronic band structure}\label{sec:band_structure}

We now turn to an analysis of the dispersion of quasiparticle bands in $\iridate$. This requires to analytically
continue our imaginary-frequency data to the real axis. We use Pad\'e
approximants~\cite{vidberg_pade_1977} to find
$\tilde{\Sigma}^\mathrm{latt}(\omega,\bk)$ from the knowledge of the periodized
lattice self-energy $\tilde{\Sigma}^\mathrm{latt}(i\omega_n, \bk)$. The
resulting band structure is shown in Fig.~\ref{bands_article} where we compare
the insulating state at $\mu = 0.8$eV (left panels) and the electron doped state
at $\mu = 1.36$eV, $\delta_{\text{per}}=7\%$, $\delta_{\text{cluster}}=10\%$
(right panels).  On the upper panels, we show the non-interacting bands
obtained by diagonalizing the TB + SO Hamiltonian $H^\mathrm{eff}_{1/2}$
(dashed lines) and the quasiparticle bands obtained from the solutions of
\begin{equation}\label{qp_eq}
 \det \Big\{ (\omega + \mu) \mathbf{1} - H^\mathrm{eff}_{1/2}(\bk)
  - \tilde{\Sigma}^\mathrm{latt}(\omega, \bk) \Big\} = 0.
\end{equation}
Bands are plotted along the $(\pi,\pi)$ - $(\pi/2,\pi/2)$ - $(0,0)$ -
$(\pi,0)$ - $(2\pi,0)$ path of the full Brillouin zone. Lower panels display the
corresponding total spectral intensity $\mathrm{Tr} \hat{A}(\omega, \bk)$.

In the insulating region, the Mott gap is clearly visible. The band structure
indicates that correlation effects have split the original non-interacting
bands. This is compatible with the observation that, at $\mu=0.8$eV, the cluster
self-energies take very different values $\mathrm{Re}\Sigma_+(0) \ne
\mathrm{Re} \Sigma_-(0)$. Lifetime effects are also not very strong and the
bands are fairly coherent, consistent with the fact that $\mathrm{Im}
\Sigma_\pm(0) \simeq 0$. The top of the lower band is located at $\simeq -0.4$
eV at the node and at $\simeq -0.2$ eV at the antinode. There is a direct gap
to the unoccupied states of the order of 0.8 eV at $\bk = (\pi/2,\pi/2)$, while
the smallest overall gap is indirect and of order 0.6 eV. Note that the latter
value is consistent with the width of the Mott plateau in
Fig.~\ref{Densities_beta200_article}.

As we move to the doped region, the Mott gap first closes at the nodal point
$\bk = (\pi/2,\pi/2)$ and the quasiparticle bands merge. The crossing of the
upper band at two points close to $(\pi/2,\pi/2)$ is a signature of the pocket
seen in the previous spectral intensities. Around these points, a clear
renormalization of the Fermi velocities by a factor $1/4$ is visible as compared to the
non-interacting bands. For $\mu = 1.36$eV there
is still a gap between the bands at $\bk = (\pi,0)$ but the lower band just
reaches the Fermi level yielding some antinodal spectral weight. It is
interesting to note that the correlation effects are much stronger on the lower
band than on the upper band.  Quasiparticles are then better defined at
$(\pi/2,\pi/2)$ (they correspond to a crossing of the upper band) than at
$(\pi,0)$ where they are associated with the lower band. This is explained by
the fact that the physics of the lower band is mainly controlled by the cluster
$\Sigma_+$, while the upper band is controlled by $\Sigma_-$. As a result, the
larger negative imaginary part of $\Sigma_+$ (see
Fig.~\ref{Extrapolation_Sigma_beta200_article}) induces stronger lifetime
effects at the antinode, while the smaller imaginary part of $\Sigma_-$
maintains coherent quasiparticles at the node.

\begin{figure}
\centering
\includegraphics[width=0.49\textwidth]{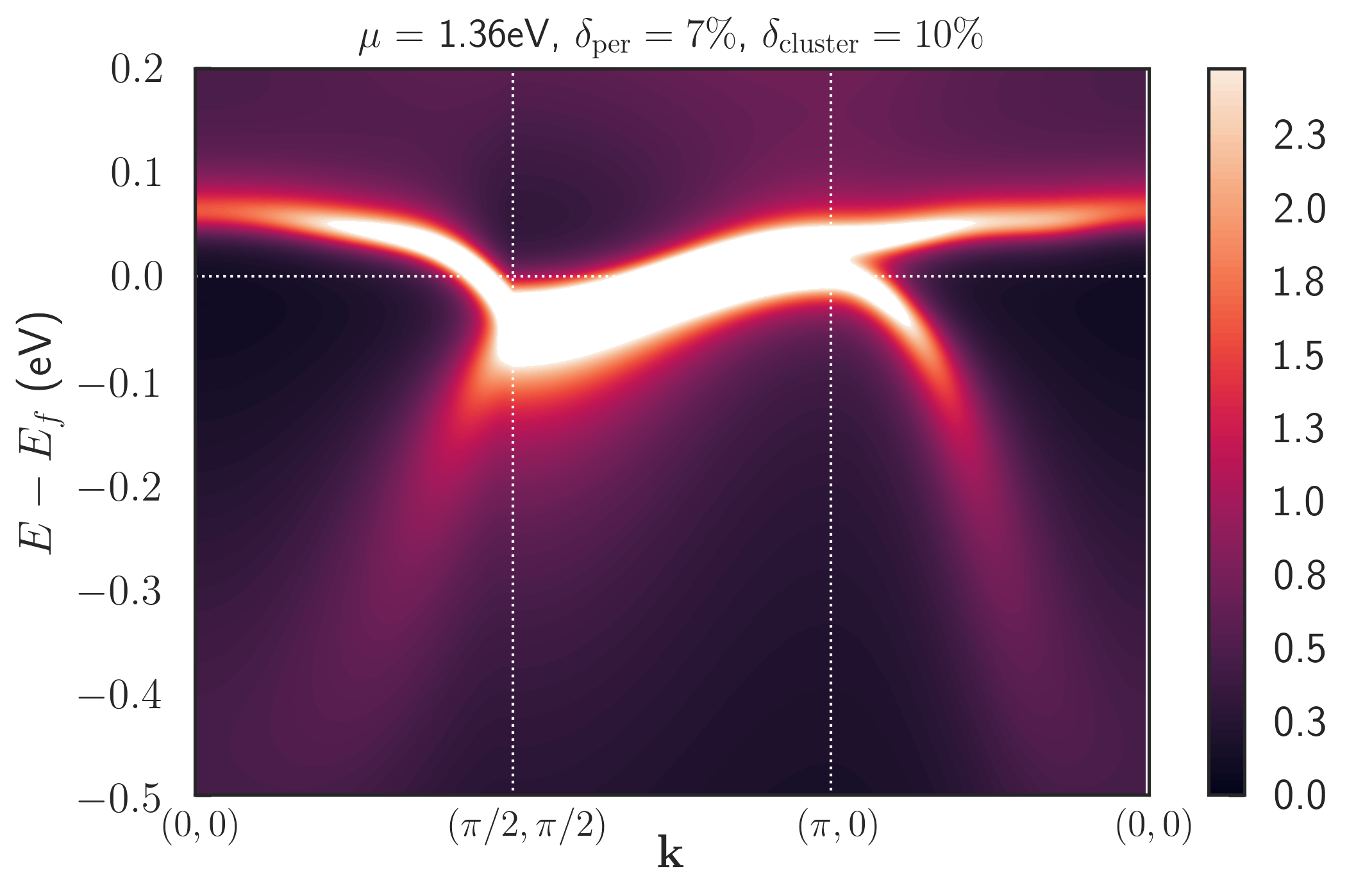}
\caption{Spectral intensity of the doped compound ($\mu=1.36$eV) along
the degenerate path $(0,0)$ - $(\pi/2,\pi/2)$ - $(\pi,0)$ - $(0,0)$ in
the full Brillouin zone. Periodized
self-energies are analytically continued from Pad\'e approximants.
$U=2$ eV, $T=58$ K.}
\label{Intensity_degenerate_path}
\end{figure}

We finally display in Fig.~\ref{Intensity_degenerate_path} a spectral intensity map along the 
$(\pi/2,\pi,2)$-$(\pi,0)$ Brillouin zone path, which corresponds to the path along which the non-interacting bands are 
degenerate. ARPES data along this path have not appeared in print to our knowledge, and our results could be useful 
in the context of future analysis of ARPES experiments.

%
%

\section{Discussion and conclusions}\label{sec:conclusion}

Finally, we discuss the comparison of our results with ARPES and other experiments on doped $\iridate$. 

Overall, there is excellent qualitative agreement. 
Comparing panels (a) and (b) of Fig.~\ref{bands_article}, a 
clear `collapse' of the Mott gap is found upon doping the insulator (i.e the two bands become much closer to 
each other). This effect was reported in ARPES experiments:~\cite{geneva_prl_2015,brouet_prb_2015}
It is clearly apparent for example in Fig.2 (g-h) of Ref.~\onlinecite{geneva_prl_2015} in which the top of the 
band at $(\pi/2,\pi/2)$ moves from about $-0.4$~eV to about $-0.1$~eV (band crossing) upon doping.  
In fact, the location of the top of the band at the `node' ($-0.4$~eV) and `antinode' (~$-0.2$~eV) 
are in good quantitative agreement with the values reported in Ref.~\onlinecite{geneva_prl_2015}. 
The rather round and spread behavior of the band at the node quite agrees with 
the experiments even if the nodal part does not appear to be as narrow as it is observed. 

The `nodal-antinodal' differentiation and formation of a pseudogap near the `antinode' is also 
consistent with the experimental observations.~\cite{geneva_prl_2015,kim_science_2014} Here, we have shown 
that the physical origin of the pseudogap is indeed the same than in cuprate superconductors, 
namely short-range spin correlations (see e.g. Refs.~\onlinecite{gunn_prl_2015,wu_prb_2017} for recent theoretical studies). 

The value of the interaction parameter $U=2$~eV for which we chose to perform our calculations should also be discussed 
in the context of experimental measurements, especially of experimental determinations of the Mott gap. With this value, we 
find a Mott gap which is indirect and of order $\sim 0.6$~eV - corresponding to the transition between the top of the 
lower Hubbard band at $(\pi,0)$ and the bottom of the upper Hubbard band at $(\pi/2,\pi/2)$ in Fig.~\ref{bands_article}(a), and 
also to the width of the Mott plateau in Fig.~\ref{Densities_beta200_article}.  
The value of the optical gap would be slightly larger. In Ref.~\onlinecite{brouet_prb_2015}, $\iridate$ was studied under both 
hole (Rh) and electron (La) doping, allowing for a determination of a Mott gap of order $~0.7$~eV, in rather good agreement with the 
present work. 
Optical spectroscopy measurements (see e.g Fig.4 in Ref.~\onlinecite{kim_prl_2008}) do reveal a sharp increase of 
absorption in that frequency range, but a rather slow onset of the optical conductivity is observed with spectral weight below 
this scale, possibly suggesting a significantly smaller value of the actual gap (although a precise determination is difficult). 
This suggests that the value of $U$ used in the present work may be a bit too large. 
Accordingly, we note that the Fermi surface renormalizations  obtained above appear to be somewhat larger 
than the values reported in Ref.~\onlinecite{geneva_prl_2015}. 

An ab-initio determination of the screened $U$ appropriate for the low-energy
model used here, as well as a more systematic study of this model as a function
of $U$ would be desirable in future work.  In connection with the latter, a
study of the possible superconducting instability as a function of $U$ can be
performed within cluster extensions of DMFT (CDMFT or DCA) for the present
model and would shed light on the elusive superconductivity of doped
$\iridate$.


\begin{acknowledgments}
We are grateful to Felix Baumberger, Alberto de la Torre, Sara Ricco and Anna
Tamai for sharing with us their experimental results and for numerous
discussions.  We also acknowledge discussions with Luca Perfetti, V\'eronique
Brouet, Dirk van der Marel, Nimrod Bachar, Silke Biermann, and thank the CPHT
computer support team for their help.
This work has been supported by the European Research Council grant
ERC-319286-QMAC and the Swiss National Science Foundation (NCCR MARVEL).  The
Flatiron Institute is supported by the Simons Foundation. 
\end{acknowledgments}


\begin{appendix}

\section{Expression of $H_0$ in the $j$ basis}
\label{app:change_basis}

Labeling $l(\bk) = e^{-i\frac{k_x + k_y}{2}}$, we have
\begin{widetext}
\begin{equation}
H_{1/2}(\bk) = \begin{pmatrix}
\frac{1}{3} \left[ \Delta_t + e_1 \left(\frac{t_1(\bk)}{t_0}\right)^2\right] + \lambda & -\frac{8}{3}l(\bk)t_1(\bk) \\
-\frac{8}{3}l^\dagger(\bk) t_1(\bk) & \frac{1}{3} \left[ \Delta_t + e_1 \left(\frac{t_1(\bk)}{t_0}\right)^2\right] + \lambda
\end{pmatrix},
\end{equation}
and
\begin{equation}
M^\dagger(\bk) = \begin{pmatrix}
-\frac{\sqrt{2}}{3}\left[\Delta_t + e_1\left(\frac{t_1(\bk)}{t_0}\right)^2\right] & \frac{2\sqrt{2}}{3}l(\bk)t_1(\bk) \\
\frac{2\sqrt{2}}{3}l^\dagger(\bk) t_1(\bk) & -\frac{\sqrt{2}}{3}\left[\Delta_t + e_1\left(\frac{t_1(\bk)}{t_0}\right)^2\right] \\
0 & \frac{2}{\sqrt{6}} l(\bk) (t_2(\bk) - t_3(\bk)) \\
\frac{2}{\sqrt{6}}l^\dagger(\bk) (t_2(\bk)-t_3(\bk)) & 0
\end{pmatrix},
\end{equation}
and
\begin{equation}
H_{3/2}(\bk) = \begin{pmatrix}
\frac{2}{3} \left[ \Delta_t + e_1 \left(\frac{t_1(\bk)}{t_0}\right)^2\right] - \frac{\lambda}{2} & -\frac{10}{3}l(\bk)t_1(\bk) & 0 & \frac{l(\bk)}{\sqrt{3}}(t_2(\bk)-t_3(\bk)) \\
-\frac{10}{3}l^\dagger(\bk) t_1(\bk) & \frac{2}{3} \left[ \Delta_t + e_1 \left(\frac{t_1(\bk)}{t_0}\right)^2\right] - \frac{\lambda}{2} & \frac{l^\dagger(\bk)}{\sqrt{3}}(t_2(\bk)-t_3(\bk)) & 0 \\
0 & \frac{l(\bk)}{\sqrt{3}}(t_2(\bk)-t_3(\bk)) & -\frac{\lambda}{2} & -2l(\bk)t_1(\bk) \\
\frac{l^\dagger(\bk)}{\sqrt{3}}(t_2(\bk)-t_3(\bk)) & 0 & -2l^\dagger(\bk) t_1(\bk) & -\frac{\lambda}{2}
\end{pmatrix}.
\end{equation}
\end{widetext}

\section{Green's functions and self-energy in the four doping regimes}
\label{app:matsubara}

\begin{figure*}
\centering
\includegraphics[width=\textwidth]{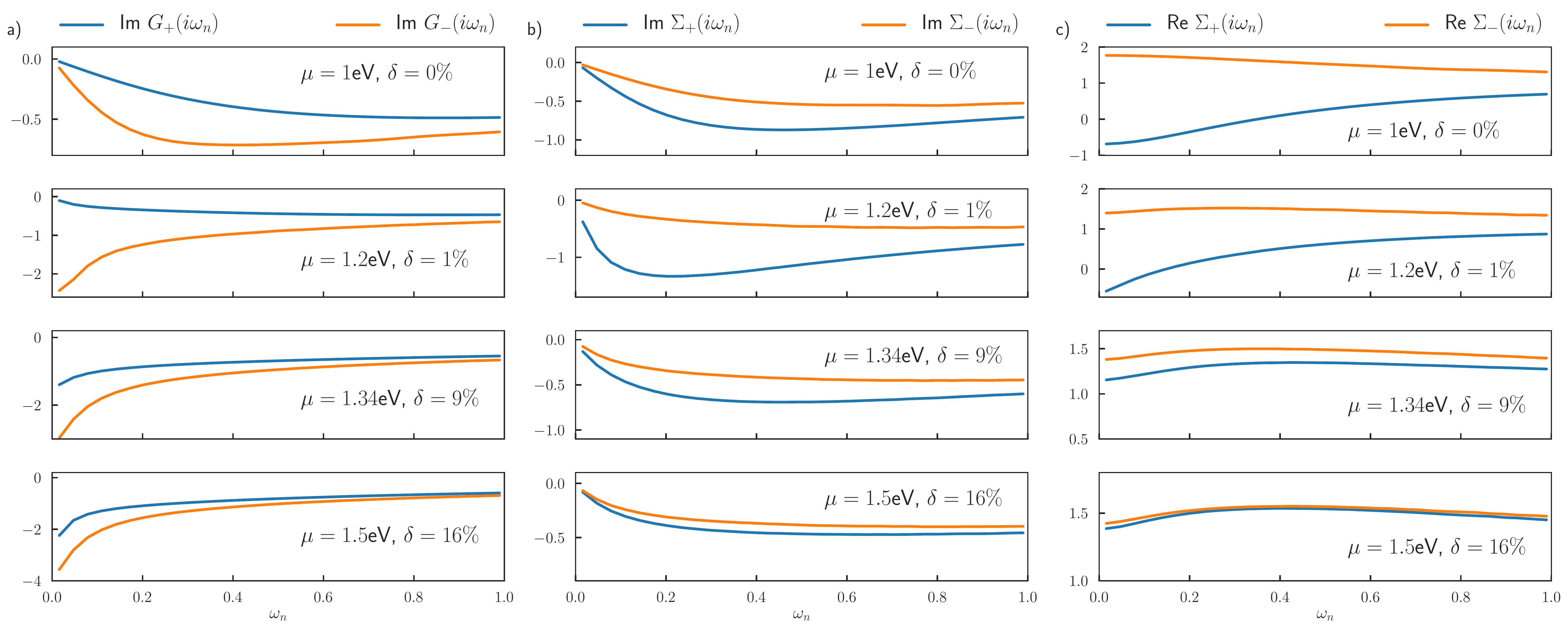}
\caption{Evolution in the even - odd basis of a) the imaginary part
of the Green's function $G_\pm(i\omega_n)$, b) the real part of the self-energy
$\Sigma_\pm(i\omega_n)$, c) the imaginary part of $\Sigma_\pm(i\omega_n)$ as a
function of the Matsubara frequency $\omega_n$. On all plots, the even (odd)
contribution is plotted in blue (orange). All quantities are depicted for
different values of the chemical potential: the upper panel corresponds to $\mu
= 1$eV and an electron doping $\delta=0\%$, below $\mu=1.2$eV, $\delta=1\%$,
then $\mu=1.34$eV, $\delta=9\%$ and finally $\mu=1.5$eV, $\delta=1.6\%$. Results
were obtained for $U=2$ eV, $T=54$K.}
\label{Cluster_G_beta200_article}
\end{figure*}

In Fig.~\ref{Cluster_G_beta200_article}, we show the Matsubara frequency
Green's functions and self-energies in the four doping regimes discussed in the  main text.
These regimes are here associated with four different values of the chemical potential corresponding to the four
rows of the figure.

For $\mu = 1$eV the system is a Mott insulator, as can be seen from the insulating character of the two components of the Green's function $G_\pm$.
Let us note that the real parts of the self-energies are very different, which is responsible for the opening of the Mott gap (see main text).
Increasing the doping, we enter a pseudogap phase.  At $\mu = 1.2$eV, the even component of the Green's function has an insulating behavior while
the odd one is metallic. At $\mu =1.34$eV, the system is in a differentiate regime. Both components of
the Green's function are now metallic but the self-energies are still quite differentiated. Going to even larger dopings we
finally reach the uniform Fermi liquid state. Hence at $\mu = 1.5$eV, we see that $G_+$ and $G_-$ are both metallic and that
the self-energies tend to be identical.

\section{Solving CDMFT equations}
\label{app:cdmft}

In order to solve the CDMFT equations, it is convenient to work in the $\pm$
basis introduced in Eq.~\eqref{eq_pm}. In this basis, the lattice Green's function is
\begin{equation}
  \hat{G}^\mathrm{latt}_\pm(i\omega_n,\bk) =
    \Big\{ (i\omega_n + \mu) \mathbf{1} - H^\mathrm{eff}_\pm(\bk)
    - \hat{\Sigma}_\pm(i\omega_n) \Big\}^{-1},
\end{equation}
where $H^\mathrm{eff}_\pm(\bk)$ is the effective $\jeff=1/2$
Hamiltonian expressed in the $\pm$ basis and the
cluster self-energy is diagonal because $A$ and $B$ sites are electronically equivalent
\begin{equation}
  \hat{\Sigma}_\pm(i\omega_n) = \begin{pmatrix} \Sigma_+(i\omega_n) & 0 \\
  0 & \Sigma_-(i\omega_n) \end{pmatrix}.
\end{equation}
Note that for a given $\bk$ point, $\hat{G}^\mathrm{latt}_\pm(i\omega_n, \bk)$
is not diagonal. One can however show that, for a generic $2 \times 2$ diagonal
matrix $\mathcal{M}$,
\begin{equation}
  \sum_{\bk \in \mathrm{RBZ}} \left[ H^\mathrm{eff}_\pm(\bk) + \mathcal{M} \right]^{-1}
\end{equation}
is a diagonal matrix too. As a result, the CDMFT self-consistency
Eq.~\eqref{eq:cdmft} becomes diagonal and reads
\begin{equation}
  \hat{G}_\pm (i\omega_n) = \sum_{\bk \in \mathrm{RBZ}} \Big\{ (i\omega_n + \mu) \mathbf{1}
  - H^\mathrm{eff}_\pm(\bk) - \hat{\Sigma}_\pm (i\omega_n) \Big\}^{-1},
\end{equation}
where both cluster quantities $\hat{G}_\pm$ and $\hat{\Sigma}_\pm$ are
diagonal. This equation is solved iteratively in the following way:
At the iteration step $n$, the quantum impurity model is described
by a non-interacting Green's function $G^{(n)}_{0, \pm}$ and a local
interaction Hamiltonian that has the following expression in the
$\pm$ basis
\begin{equation}
  \begin{split}
  \mathcal{H}^\mathrm{int} = \frac{U}{2} \sum_{s = \pm} & \left( n_{s\uparrow}n_{s\downarrow} + n_{s\uparrow}n_{\bar{s}\downarrow} +  \right. \\
 & \left. c_{s\uparrow}^\dagger c_{s\downarrow}^{\dagger}c_{\bar{s}\downarrow}c_{\bar{s}\uparrow} + c_{s\uparrow}^\dagger c_{\bar{s}\downarrow}^\dagger c_{s\downarrow}c_{\bar{s}\uparrow} \right).
  \end{split}
\end{equation}
This cluster model is solved using the CT-HYB quantum impurity solver. This
solver directly works in the $\pm$ basis. It yields both the cluster Green's
functions $G_\pm^{(n)}$ and self-energies $\Sigma^{(n)}_\pm$. The self-consistency
condition is used to construct a local diagonal lattice Green's function 
\begin{equation}
  \hat{G}^{(n)}_{\mathrm{loc}, \pm} (i\omega_n) = \sum_{\bk \in \mathrm{RBZ}} \Big\{ (i\omega_n + \mu) \mathbf{1}
  - H^\mathrm{eff}_\pm(\bk) - \hat{\Sigma}^{(n)}_\pm (i\omega_n) \Big\}^{-1}.
\end{equation}
This allows to get a new expression for the non-interacting cluster
Green's function, via a modified Dyson equation:
\begin{equation}
  \left[ G^{(n+1)}_{0, \pm} \right]^{-1} = \left[ G^{(n)}_{\mathrm{loc}, \pm} \right]^{-1} + \Sigma^{(n)}_\pm.
\end{equation}
This procedure is iterated until convergence.

\section{Absence of a Mott insulator with the standard periodization scheme}
\label{app:periodization}

The usual periodization of the self-energy writes
\begin{equation}
  \tilde{\Sigma}^\mathrm{latt}(i\omega_n, \bk) = \begin{pmatrix}
  \Sigma_{AA} & \Sigma_{AB} \times f(\bk) \\
  \Sigma_{AB} \times f^*(\bk) & \Sigma_{AA}
  \end{pmatrix},
\end{equation}
where
\begin{subequations}
\begin{align}
f(\bk) & = \frac{1}{4}\left(1+e^{-ik_x}+e^{-ik_y}+e^{-i(k_x+k_y)}\right) \\
& = \cos \frac{k_x}{2} \cos \frac{k_y}{2}e^{-i\frac{k_x+k_y}{2}}.
\end{align}
\end{subequations}
$\bk = (k_1,k_2)$ is expressed in the reduced Brillouin zone.
We see from Fig.~\ref{tb_model} that the degeneracy of the $(\pi/2,\pi/2)$ - $(\pi,0)$ path
in the full Brillouin zone has to be lifted in order to create a Mott insulating gap.
However $f(\bk) = 0$ along this path and the self-energy has the following
expression
\begin{equation}
  \tilde{\Sigma}^\mathrm{latt}(i\omega_n, \bk) = \Sigma_{AA} \times \mathbf{1}_{2\times2}.
\end{equation}
Hence the self-energy only renormalizes the chemical potential in the
quasiparticle equation (Eq.~\eqref{qp_eq}) at $\omega=0$, forbidding any
lifting of the degeneracy between the quasiparticle bands and therefore any gap
in the band structure.

\end{appendix}

\bibliography{main}

\end{document}